\documentclass[11pt]{article}
\usepackage[utf8]{inputenc}
\usepackage{amsmath}
\usepackage{amsthm}
\usepackage{amsfonts}
\usepackage{natbib}
\usepackage{bm}
\usepackage{thm-restate}

\usepackage{graphicx}
\usepackage{subcaption}
\usepackage{float}

\usepackage{color}   
\usepackage{hyperref}
\hypersetup{
    linktoc=all,     
    linkcolor=blue,  
}

\usepackage{framed}
\usepackage{setspace}
\usepackage{listings}
\spacing{1.06}
\usepackage[margin=1in]{geometry}

\usepackage{todonotes}

\usepackage[linesnumbered, ruled,vlined]{algorithm2e}

\theoremstyle{plain}

\theoremstyle{definition}

\theoremstyle{remark}

\usepackage{dsfont}

\usepackage{todonotes}

\usepackage{authblk}

\title{NaijaCoder: Participatory Design for Early Algorithms Education in the Global South}

\author[1]{
Daniel G. Alabi
}
\affil[1]{
Columbia University
}

\author[2]{
Atinuke	Adegbile
}
\affil[2]{
Global Integrated Education Volunteers Association (GIEVA)
}

\author[3]{
Lekan Afuye
}
\affil[3]{
Cornell University
}

\author[4]{
Philip	Abel
}
\affil[4]{
Twilio, Inc.
}

\author[5]{
Alida Monaco
}
\affil[5]{
ICF International
}

\date{}

\begin{document}

\maketitle

\begin{abstract}

The majority of Nigerian high schoolers have little to no exposure to the basics of algorithms and programming. We believe this trajectory should change as programming offers these students, especially those from indigent backgrounds, an opportunity to learn profitable skills and ignite their passions for problem-solving and critical thinking.

NaijaCoder is an organization that is dedicated to organizing a free, intensive summer program in Nigeria to teach the basics of algorithms and computer programming to high schoolers. However, the adoption of computer science curriculum has been especially challenging in countries in the global south that face unique challenges---such as unstable power supply, internet service, and price volatility. We design a curriculum that is more conducive to the local environment while incorporating rigorous thinking and preparation. Using basic survey designs, we
elicit feedback, from the students, designed to further improve and iterate on our curriculum.

\end{abstract}

\section{Introduction}

``The global south'' is a term that is used to refer to countries and regions located in the southern hemisphere, primarily in Africa, Latin America, Asia, and Oceania. These countries are often characterized by lower levels of economic development and challenges related to issues such as poverty, inequality, and limited access to resources and opportunities~\citep{GlobalSouth}. 

While the majority of the initial literature on issues faced
by the global south is focused on negative aspects of the region, we are now seeing an emergence of literature on self-sufficiency, growth, self-actualization, and \textbf{talent development}~\citep{dargin2013rise}.

In this paper, we examine a new talent development initiative developed by NaijaCoder.
Most high school students in the global south lack significant exposure to fundamental algorithms and programming concepts. NaijaCoder is of the opinion that this direction must shift, as programming presents these students, particularly those facing financial challenges, with a chance to acquire valuable skills and spark their interests in creative problem-solving and essential critical thinking.
NaijaCoder is a non-profit organization that was founded with one major goal in mind: to proliferate computer science education throughout Nigeria. To achieve this goal, NaijaCoder will target early-stage development of technical, critical-thinking, and communications skills from the secondary school level. 

The future of Nigeria lies in its youth. We hope to further strengthen the outlook of the country, especially in the northern parts, by continuing
the execution of the program.
At a high level, we want the enrolled student to gain significant experience in computer programming. Given the ever-increasing need for computer and data science skills, we believe equipping the students with computer science skills will further diversify the technical skills of Nigerian students. African and Black students are severely underrepresented in STEM fields around the world. 
One of the reasons for this under-representation
potentially stems from a lack of resources and role models~\citep{TC07, goff2014essence, wright2018brilliance, WOGB22}. We hope that our program can remedy this situation by providing free instruction for in-demand STEM skills. In terms of long-term sustainability, we plan to run the program every year and will revise the syllabus to incorporate major trends in computer programming. We will also expand the class offerings to include multiple levels of classes for repeat students, and include hackathons and mentorship circles. 

Youth in Nigeria (especially in the northern parts of Nigeria) have little exposure to computer programming. This lack of exposure hinders future opportunities (e.g., for being hired to work on software products or starting a technical company). NaijaCoder is geared towards using computer science as a tool for a better economic, social, and academic life for youth. Given the diversity of the occupations and interests of the board members, we can provide mentorship on various career options — both of the technical and non-technical variety. Our team consists of Nigerians who currently live and work in the USA but all attended high school in Nigeria. The occupations of the teaching staff and board members include engineers, computer science (PhDs), and business leaders. As such, we will expose the student participants to more potential career options that employ computer programming skills.

\paragraph{History}

The NaijaCoder organization was founded in 2022 and 
focuses on providing a high quality 2-week summer program in Nigeria each year. The program was founded by Nigerians who studied in the United States after high school. Administrative activities take place predominantly during the spring before each summer program.

\paragraph{Intensive 2-week Program}

We organize an annual intensive 2-week program to prepare the students for future computer science exploration. By the end of the program, we hope to accomplish the following goals:
\begin{enumerate}
\item Every student should be familiar with every topic in the syllabus. The topics span themes in basic data structures and introductory computer science. We have daily quizzes on the curriculum taught each day to gauge understanding.
\item The students should gain skills in collaborating with each other to solve programming problems. We measure progress by how comfortable they are talking to their peers and how well they work together to produce correct results.
\item Students should develop a systematic approach to problem solving. We measure progress by how well they identify tools (e.g., dynamic programming or tree data structure) needed to solve problems.
\end{enumerate}

\paragraph{Strengthening U.S.-Nigeria ties}

Through NaijaCoder programs, we hope to further strengthen ties between the U.S. and Nigeria by leveraging our connections in both countries. Given that the majority of the key personnel obtained undergraduate and graduate degrees in the U.S, we share our positive experiences with the secondary school students and encourage them to apply to U.S. colleges or other colleges abroad. For example, we had a general question and answer session on the final day of the program in 2023 to talk through the teachers' experiences studying in the U.S., and how Nigerian students can apply to educational programs abroad. We additionally had two guest lecturers from the Global Integrated Education Volunteers Association (GIEVA) and EducationUSA office from the U.S. Embassy in Abuja who spoke about student resources for applying to colleges abroad.

\section{Prior and Related Work}

In this section, we discuss some previous work that informs our evaluations and study. First, we examine and highlight existing centers for STEM (Science, Technology, Engineering, Mathematics) development in Nigeria. A large fraction of STEM outreach efforts is focused on college-level students while NaijaCoder efforts are targeted towards pre-college/K-12 students and educators. Next, a crucial challenge to the Nigerian education system is the unavailability of basic
infrastructure (such as constant power supply). So any curriculum must be designed to work around some of such challenges. In particular, \textit{all exercises cannot be designed for the computer lab alone!} Exercises on paper are equally important, as the instability of the power supply requires a flexible curriculum. We also compare our work to other programs in the global south designed for educating high school students. A prominent example is AddisCoder
~\citep{addiscoder}. Lastly, we examine the existing literature on education methodology and curriculum design to guide our syllabus elements.

\subsection{Existing Centers for STEM in Nigeria}

Before NaijaCoder was founded, there already existed myriad initiatives to encourage science and engineering. We briefly discuss some of such initiatives. However, we stress that these initiatives are focused on the \textbf{post-graduate levels}!

\paragraph{Nigerian Mathematical Society}

The Nigerian Mathematical Society (NMS) is a professional association dedicated to promoting the study, research, and application of mathematics in Nigeria. It serves as a platform for mathematicians, educators, researchers, and students to collaborate, exchange ideas, and advance the field of mathematics in the country.
Most of their initiatives, thus far, have been focused on raising public awareness about the importance of mathematics in various aspects of society~\citep{nms}. The NMS organizes conferences, seminars, workshops, and lectures to facilitate the exchange of mathematical knowledge and ideas.
Members of the society include mathematicians, educators, researchers, students, and enthusiasts who share an interest in mathematics. 

However, the majority of activities organized by the Nigerian Mathematical Society is geared toward college-level educators and not high-school or K-12 students or educators.

\paragraph{National Mathematical Centre}

The center is currently focused on promoting mathematical activities within colleges and secondary schools in Abuja, Nigeria. During the semesters, the center organizes training programs, workshops, and research activities to enhance mathematical education and research in the country~\citep{nmc}.

\paragraph{National Space Research and Development Agency (NASRDA)}

NASRDA is responsible for space research and technology development in Nigeria. It engages in activities related to satellite development, space science research, and applications for various sectors including agriculture, communications, and weather forecasting. Although NASRDA promotes STEM activities in Nigeria, it is primarily engaged in research, development, and often political activities.

We also wish to highlight the Centre For Space Transport and Propulsion (CSTP).
It is one of the activity centres of the Nigerian National Space Research and Development Agency (NASRDA). Currently located at the Federal University of Technology, Akure, CSTP focuses on research and development in space propulsion and related technologies.

Another activity center for NASRDA is the Center for Atmospheric Research (CAR). It is a National Space Research and Development Sub-Agency. CAR conducts research in atmospheric science, climate studies, and related fields to better understand atmospheric processes and their impacts on society.

\paragraph{African University of Science and Technology (AUST)}

AUST is a private, pan-African institution currently located in Abuja. It offers graduate and postgraduate programs in various STEM disciplines, including computer science, materials science, and petroleum engineering.

\paragraph{National Agency for Science and Engineering Infrastructure (NASENI)}

NASENI is an agency under the Nigerian Ministry of Science and Technology that aims to promote the development and commercialization of indigenous technologies. It operates research institutes in various engineering and technological fields.

\paragraph{Federal Institute of Industrial Research, Oshodi (FIIRO)}

FIIRO is a government-owned research institute that works on industrial research and development, including food processing, chemicals, and materials science.

\subsection{Infrastructure Challenges}

Most computer programming courses require access to at least one computer. Traditional computers require semi-regular access to electricity to function. 
If a student or instructor does not have access to
a computer and/or electricity, this can
impede the learning environment~\citep{poverty_brief}. We
now review some infrastructure challenges in
Nigeria.

\paragraph{Electricity and Power Supply}

Nigeria struggles with providing a consistent and reliable power supply~\citep{IBITOYE2007492, ABSD19}. Frequent power outages, inadequate generation capacity, and an outdated power infrastructure have, unfortunately,
hindered economic growth and industrial development~\citep{IYKE2015166}.

In designing the NaijaCoder 2023 syllabus, the
instructors \textit{always} had a daily
back-up plan, in case there was a power outage.
To give a specific example, if the current
instructor was demonstrating bubble sort using an
electrical projector, then the instructor would 
switch to the use of a projector connected to a
``power bank.''

Power outages were experienced regularly during the NaijaCoder 2023 program, usually occurring at least once a day for as little to 10 minutes to at most an entire day. 

\paragraph{Transportation}

The transportation sector in Nigeria has faced challenges in terms of inadequate public transportation systems and congestion on roads. These issues have hindered efficient movement of goods and people (including students).

The NaijaCoder 2023 team situated the program in
the geographic
center of Nigeria (Abuja) which also has the
most urban transportation infrastructure in the country.
In subsequent years, the team hopes to provide
further transportation assistance to the students in the form of subsidies.

\paragraph{Internet Connectivity}

While internet access has improved, there are still disparities in internet penetration between urban and rural areas. Limited access to reliable and affordable internet can impact digital literacy and economic opportunities.
The school where the 2023 program was held had
semi-constant internet availability. During electricity power outages, however, the school would lose internet connectivity. 

\subsection{Other Computer Science Programs for High School Students within Africa}

The NaijaCoder initiative was especially inspired
by the AddisCoder program, an initiative that
teaches Ethiopian high-school students how to code.
However, because AddisCoder is implemented in a different developing context within Ethiopia, the curriculum and methodology itself is not directly translatable to Nigeria ~\citep{addiscoder}~\citep{NaijaCoder}.
For example, AddisCoder (as of 2023) is a
4-week summer program while NaijaCoder is a 2-week program.
Also, AddisCoder (as of 2023) does not
have guest lectures about applying to colleges or about
adjacent fields to computer science.
So the time-period for instruction during the NaijaCoder summer
camp is shorter while still incorporating
rigorous preparation.

Ethiopia has a smaller population than Nigeria,
is in the Eastern part of Africa, has a
different high-school curricula, and has a
high-school calendar that is
\textbf{significantly} different from Nigeria's.
This governs why NaijaCoder's syllabus and
calendar of instruction has to be greatly
adjusted to reflect the differences in country
policies, politics, and education 
infrastructure.

Within West Africa (where Nigeria is located),
the most recognized standardized test is the
West African Senior School Certificate Examination (WASSCE).
Computer science (or strongly related courses) is not 
an approved subject~\citep{wassce}.
As a result, schools are, unfortunately, not
incentivized to teach the basics of computation in high school. The NaijaCoder
summer camp provides a much-needed introduction to computer
science that will be relevant for future studies at the
university level. For the 2023 camp, we partnered with
Nigerian-based institutions, such as GIEVA (the
Global Integrated Education Volunteers Association) and plan to
partner with other Nigerian-based institutions in the future.
Additionally, we plan to, more
seriously, nationally advocate for computer science
education for high school students.

\subsection{Literature on Education Methodology}

In designing a syllabus for the
NaijaCoder 2023 summer program, we also examined
some significant previous literature on curricula
design and education (both within computer science
and beyond)~\citep{Webb17, Clear20, Taylor2018EnablingMC, Rao05, Blank_Kumar02}. Here
are some elements that we paid particularly attention to:

\paragraph{Constructivism}

This theory emphasizes active learning, where students construct their own understanding through experiences and interactions with the environment (``environment'' includes students, instructors, and the material). Prominent theorists include Jean Piaget and Lev Vygotsky~\citep{wilson_constructivist_2004}.
In~\citep{wilson_constructivist_2004},
the author explores the principles of constructivism and its implications for instructional design~\citep{wilson_constructivist_2004}.

For NaijaCoder, we incorporated some degree of
constructivism
by requiring that every lesson plan is
accompanied by exercises that the students
can work collaboratively or individually on.

\paragraph{Bloom's Taxonomy}

This classification system categorizes cognitive skills into a hierarchy, from lower-order thinking (remembering, understanding) to higher-order thinking (analyzing, evaluating, creating)
~\citep{bloom1956taxonomy}.

For the NaijaCoder 2023 syllabus, we 
de-emphasized memorization of terms and
focused on active learning of concepts via
examples. For example, the final quiz was open book so that students focused on understanding and applying the material rather than memorizing and regurgitating previous examples.

\paragraph{Backward Design}

This approach starts with defining desired learning outcomes and then designing instruction to achieve those outcomes. It was popularized by Grant Wiggins and Jay McTighe~\citep{wiggins2005understanding}.

For the NaijaCoder 2023 summer program, we
first emphasized learning outcomes such as verifying the correctness of algorithms such
as sorting and search algorithms. Then specific
algorithms, such as bubblesort and insertion sort,
were analyzed in great detail later on in the
syllabus.
We focused on such basic searching and sorting
algorithms because most of the students had no previous experience
with algorithms. However, for future iterations, the
syllabus will include more advanced materials.

\paragraph{Experiential Learning}

This approach focuses on learning through direct experiences, reflection, and application. David Kolb's Experiential Learning Cycle is a foundational model and his papers 
 outline the Experiential Learning Cycle and its application in educational contexts~\citep{Kolb84}.

At the beginning of every day, a NaijaCoder
instructor first reflected on the examples and
algorithms learned in the previous day and reviewed any student questions. We observed
that this exercise further reinforced the
students' understanding of the material.

\section{Syllabus}

The syllabus for the NaijaCoder 2023 summer
program was developed after several rounds of
feedback and extensive literature review. 
The feedback was provided primarily by high-school principals 
and teachers in both Nigeria and the U.S. We also received
feedback from other members of the NaijaCoder board.
The feedback focused on how much technical material should
be included in the syllabus while accommodating the needs of
students who had no previous programming experience.
This activity was performed by
NaijaCoder team members with input
from external members, as well. The syllabus was developed for a two-week time period based on the required time to teach the complex material and based on student and teacher availability.

\subsection{Syllabus Overview}
The NaijaCoder programming consists of two weeks of rigorous training in the basics of computer programming, algorithms, and relevant mathematics. The program focuses on student-led teaching, where collaboration and hands-on practice with coding and computers is the focus. Our syllabus includes the following core concepts in basic algorithms:
\begin{enumerate}
    \item Representation of numbers in binary and other forms
    \item Constructing and using a list data structure
    \item Searching algorithms (e.g., linear and binary search)
    \item Sorting algorithms (e.g., insertion and bubble sort)
    \item Constructing Recursive Algorithms (e.g., for and while loops)
      \item Computational efficiency: Big-O
\end{enumerate}
All these topics are approached via the Python programming language.

Guest lecturers were also invited throughout the program and presented on various topics including: electrical engineering and computer chip efficiencies, using programming to address climate change, and resources for students applying to universities abroad.

After the program, students understand the basics of computer programming and are able to develop simple to semi-complex functions or algorithms to perform various tasks. The entire program is taught in the python programming language, but the focus on understanding theory and logic ensures that students could easily learn other programming languages in the future. Students are also able to analyze different algorithms and estimate which one will be more efficient in terms of time and space complexity.

Future iterations of the program
will include the following targeted activities: hackathons (computing programming competitions), mentoring circles, and multiple difficulty levels of classes. 

\subsection{Program Materials}
The program loosely follows chapters 1 through 3 of
``Introduction to Algorithms''
by Thomas H. Cormen, Charles E. Leiserson, Ronald L. Rivest, and Clifford Stein~\citep{CLRS}. The term ``loosely'' is employed because the teaching is student-focused, and as such when students have questions, the teaching staff will spend time answering those questions even if they are not within the planned syllabus. This is to ensure that students' understanding of the material is holistic, and grounded in understanding rather than memorization.

The teaching staff consisted of a principle teacher, who received a PhD in theoretical computer science from Harvard University and was born and raised in Abuja, Nigeria. An additional teaching assistant was involved, and they received a bachelors from Harvard University and has studied computer science extensively. 

Guest lecturers were also invited to speak to the class. See the guest lecturers and the subjects they discussed below:
\begin{enumerate}
    \item\textit{Electrical Engineering PhD from Cornell University}: Spoke about the hardware of computers and designing efficient computer chips.
    \item\textit{Climate and Sustainability Consultant}: Spoke about using coding and algorithms to solve climate change issues.
    \item \textit{A representative from the Global Integrated Education Volunteers Association, Inc. (GIEVA)}: Spoke about GIEVA's resources for students applying to universities abroad.
    \item \textit{A representative from EducationUSA}: Spoke about EducationUSA's resources for students applying to study in the United States, and how international students should begin preparing their applications. 
\end{enumerate}
The Google Colab online coding environment was used for students to practice coding in python. All relevant program messages and announcements were sent via a slack channel that students could access on their phones or computers.

\begin{figure}[ht]
    \begin{subfigure}[b]{0.45\textwidth}
        \includegraphics[width=\textwidth]{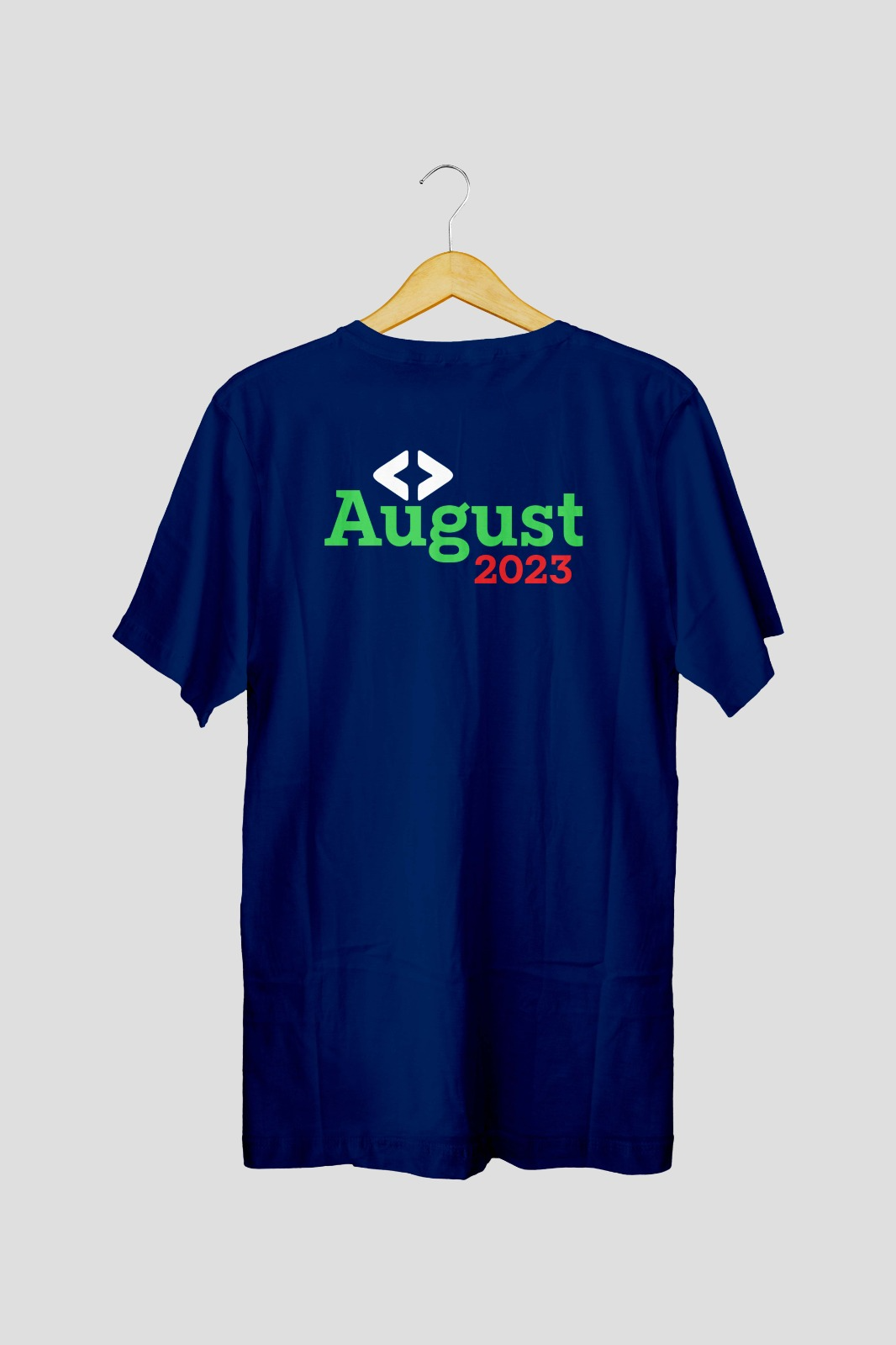}
        \label{fig:shirts1}
    \end{subfigure}
    \quad
    \begin{subfigure}[b]{0.45\textwidth}
        \includegraphics[width=\textwidth]{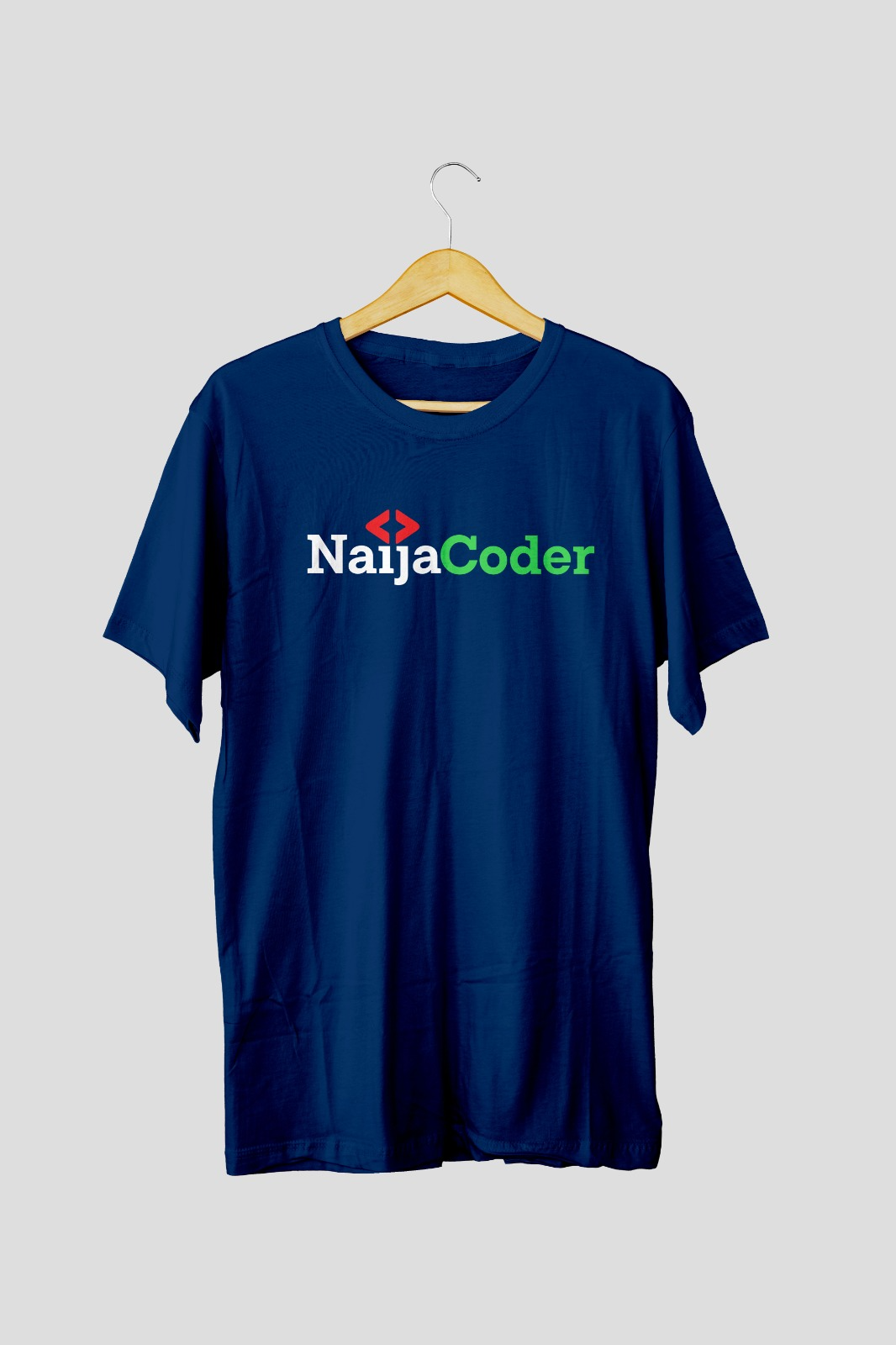}
        \label{fig:shirts2}
    \end{subfigure}
    \caption{NaijaCoder T-Shirts for Students and Instructors.}
    \label{fig:shirts}
\end{figure}

Students were also given t-shirts with the NaijaCoder logo on them. See 
Figure~\ref{fig:shirts}.

\subsection{Program Logistics}
As the program operates in a developing context, there were several infrastructural challenges that needed to be mitigated. The program provided free and semi-stable access to electricity, internet, computers and lunch every day for students. The program was held in a local private school in Abuja, Nigeria from 8 am to 3 pm Monday through Friday, with a 1 to 1.5 hour break for lunch. 

\subsection{Weekly Syllabus}
\paragraph{Week 1}
Week 1 focused on developing a deep understanding of the raw basics of computer science, starting from how computers work, to understanding different variable types, and finally to understanding the logic of computation. 

\paragraph{Week 2}
Week 2 focused on translating the students' knowledge of computer programming and logic to coding simple to semi-complex algorithms and functions in python. Week 2 also focused on quantifying the space and time complexity of various algorithms and functions. See Figure 2 below for the daily syllabus throughout the two week program.

\begin{figure*}[htp]
    \centering
    \includegraphics[width=15cm]{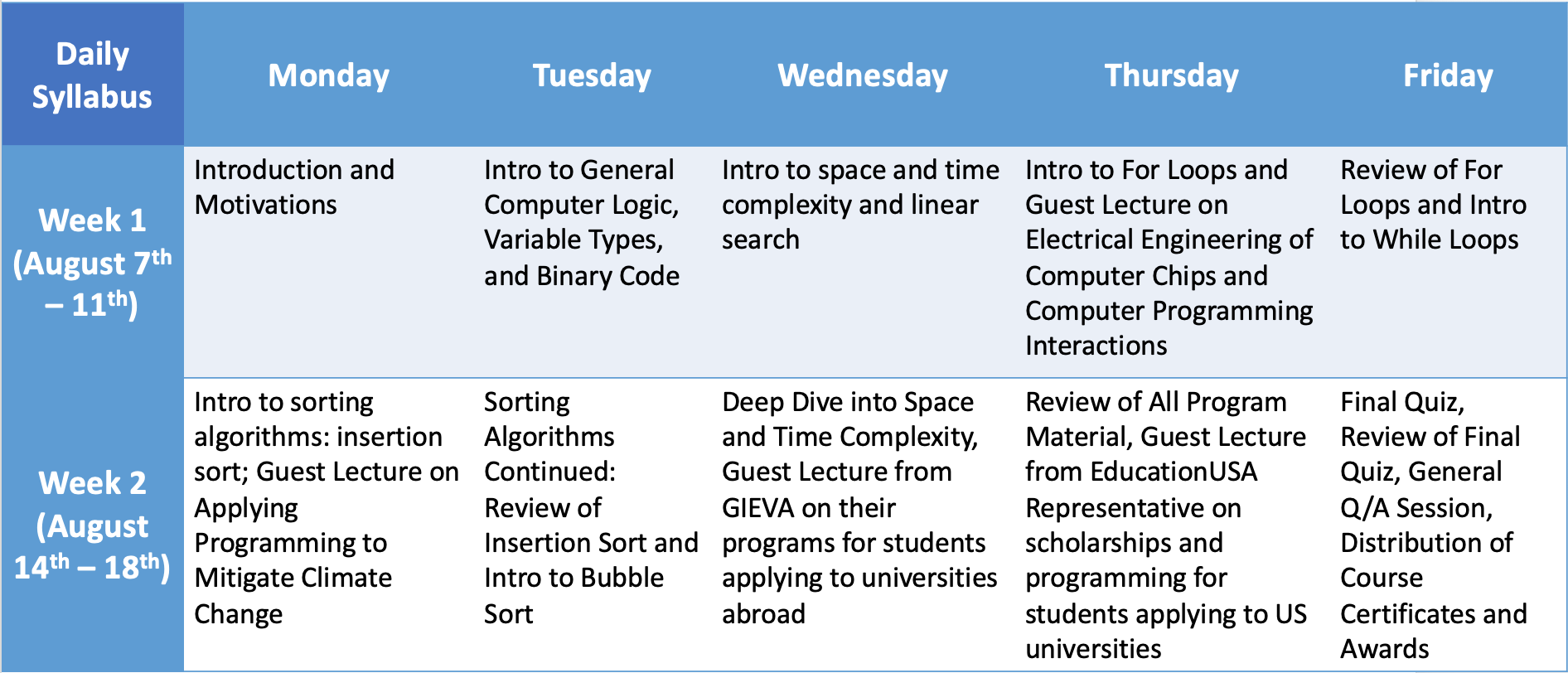}
    \caption{Daily Syllabus for NaijaCoder 2023}
    \label{tab:syllabus}
\end{figure*}

\begin{enumerate}
\item
A final quiz was provided to students so that they could measure their progress and understanding of the course material. Those who scored the highest on the exam were given cash prizes, to incentivize their continued research and study in computer programming. The teaching staff also walked through the final quiz so students could understand their mistakes. Certificates were handed out to all students upon their completion of the course. 

\item
Students were also introduced to the NaijaCoder Ambassador Program, which is for NaijaCoder alumni who are interested in volunteering to spread the word about NaijaCoder within their respective highschools. This aids in NaijaCoder recruitment for the next year, and involved alumni are able to receive recommendation letters for university applications from NaijaCoder staff. 

\end{enumerate}

\section{Data and Evidence}

Throughout the preparation and execution of
the NaijaCoder summer intensive program, we
collated data for two major reasons:
(1) To assess the effectiveness of our current
teaching strategies, and 
(2) To assess the preparedness of the cohort of
students to understand the syllabus material and
build upon the material to keep improving their 
computer science skills.

\subsection{Questions on Application Forms}

In Late 2022, the student application forms
were made public and sent to appropriate 
school authorities to distribute to their students.

Until the deadline, we received over
50 applications, some of which were
submitted online and some offline.
For the initial online forms, here are a selection
of the questions asked:
\begin{enumerate}
\item\textbf{Personal Information}: Standard biographical
questions such as first and last names, age, and
gender.
\item\textbf{School Information}: Information about
their grade levels and the school(s) from which
they plan to obtain or have obtained
their high school degrees from.
\item\textbf{State of Residence}: The state (out of
the 36 states in Nigeria, excluding the
Federal Capital Territory) where they reside.
\item\textbf{Academic and Coding History}:
Each applicant had to upload a copy of their
most recent school report card with their name legibly written on the report card.
We also asked applicants to explain any
previous computer programming experience, while
emphasizing that previous experience is not a requirement to participate in the NaijaCoder program.
There was also an optional question about whether
the applicant had any coding samples 
(via a public link) they could
share with the team.
\item\textbf{Academic Awards}:
To assess the competitiveness of the 
applicant, we also asked if they had previously
won any academic awards or competitions.
\item\textbf{Personal Essay Question}:
We also included a personal essay question,
akin to the ones they might have to write for
entry into a college in the U.S. 
This question was designed to
allow the NaijaCoder staff to
assess the resilience of the applicant. 
For example,
applicants were encouraged to chat about
their experiences with hardships and any other
relevant background.
\end{enumerate}

Eventually, the NaijaCoder staff admitted
students from a variety of schools in the
country. There were a total of 25 students
admitted, initially.
However, one particular public school stood
out with the most number of admitted students:
6 students from
Federal Government Academy in Suleja, Abuja.
The state of residence of applicants ranged
as far east as Nsukka in Enugu state and as far
north as Kano state.
The ages of the applicants ranged from
13 years old up to 29 years old.
The 29-year old was an outlier; all the other
students were less than 18 years old.

\subsection{Data from Additional Survey Questions}

Midway through the NaijaCoder program, we
sent out a survey to the (already-admitted)
students to assess their familiarity with
the syllabus and their aspirations for the program.
Some of the questions included
the following:
\begin{enumerate}
\item\textbf{Familiarity with Syllabus}:
We asked the students about how familiar they
were with the syllabus materials before NaijaCoder.
\item\textbf{Future College Aspirations}:
There were questions about how applicable
they believe
the curriculum to be for their future
college work.
\end{enumerate}

The response rate was 17.0/25 (68\%).
Out of the survey respondents,
52.9\% of the students indicated that they were
not familiar with most of the syllabus before the
program began.
The rest of the students answered ``No'' or ``Maybe''
to the question about familiarity with the syllabus.
A single student indicated that they did not think the syllabus might be useful for
their college studies.
Lastly, 100\% of the students indicated that they
would be interested in gaining more knowledge from similar programs, in the future.

\section{Conclusion \& Future Work}

Early exposure to technical computer science skills could offer students, especially those from poorer backgrounds, an opportunity to learn profitable skills and could ignite their passion for problem solving in domains beyond computer science.

NaijaCoder is a research and education institute.
The current major activity of the organization is to
organize an intensive summer program in Nigeria aimed at secondary school students (also known as high school students). The program provides free instruction on the basics of algorithms, statistics, and programming. See the website for more details~\citep{NaijaCoder}.
We launched in June of 2022 when we visited LifeGate academy and Model Secondary school in Abuja, Nigeria. We gave intermediate lecturing and connected with some students and headmasters. After realizing that demand was far more than the supply of instructors, we decided to relaunch the program in summer of 2023 with more support and structure. 
Throughout the program,
we elicited feedback from the students
and instructors to improve the curriculum
design process and outcome.



Here are some future aspects of the program
execution we will pay special attention to:
\paragraph{Emphasizing Breadth}
From the midway survey questions, a few
students indicated that they did
not think that the syllabus might be useful for
their college studies. For the next iteration of
the summer program,
we hope to emphasize (in the syllabus and 
elsewhere) the breadth of the curriculum.
In particular, the goal of the program is not
for \textit{all} students to major in
computer science but to instead \textit{equip}
the students to apply computational thinking and
skills to
any major they wish to pursue in college and
beyond.

\paragraph{Responding to Scale}

Given the interest in the in-person summer 2023
program, it is likely
that the program will grow in scale
(in terms of number of applications received and
admitted students). The NaijaCoder team needs to
respond to this challenge by also appropriately
adjusting/increasing the team,
increasing recruitment and fundraising efforts,
and expanding social media efforts (for better
channels of feedback).
One of our major focuses will be on northern Nigeria. According to the World Bank~\citep{poverty_brief}, poverty in Nigeria disproportionately affects rural households in the north of Nigeria. Among ``those living below the \$1.90 poverty line in 2018/19, 84.6 percent lived in rural areas and 76.3 percent lived in North Central, North East, or North West zones.'' 

\paragraph{Further Enrichment Activities}

Beyond the summer camp,
we plan to host further activities for the 
students to keep improving their programming
skillset.
For example, we plan on 
hosting a hackathon (fast-paced coding camps) for students to learn about creating data products using the latest software tools.
Another activity would be the implementation
of ``mentoring circles'':
given the diversity of the careers of our board members (e.g., in software development, finance, and academia), we will offer advice and mentorship on using software tools for product development.

\section{Acknowledgements}

Daniel Alabi was supported by the Simons Foundation (965342, D.A.) as part of the Junior Fellowship from the Simons Society
of Fellows.
We are grateful to the SIGCSE reviewers
for helpful feedback.

\clearpage

\bibliographystyle{ACM-Reference-Format}
\bibliography{main}


\end{document}